\documentclass[nofootinbib,twocolumn,aps,pre,superscriptaddress,citeautoscript,floatfix]{revtex4-2}
\usepackage[title]{appendix}
\usepackage{graphicx}
\usepackage{color}
\usepackage{amsmath}
\usepackage{amssymb}
\usepackage{float}
\usepackage{bm}
\usepackage[inkscapeformat=png]{svg}

\newcommand\be{\begin{equation}}
\newcommand\ee{\end{equation}}
\newcommand\bea{\begin{eqnarray}}
\newcommand\eea{\end{eqnarray}}
\newcommand{\kB }{k_{\rm B}}
\newcommand{\lB}{l_{\rm B}}

\newcommand{\phis}{\phi_{\rm s}}

\newcommand{\vecr}{\boldsymbol{r}}

\begin{document}

\title{The effects of ionic valency and size asymmetry\\ on counterion adsorption}

\author{Or Ben Yaakov}
\affiliation{School of Physics and Astronomy,
  Tel Aviv University, Tel Aviv 6997801, Israel}
\affiliation{Center for Physics and Chemistry of Living Systems,
  Tel Aviv University, Tel Aviv 6997801, Israel}

\author{Haim Diamant}
\affiliation{School of Chemistry,
  Tel Aviv University, Tel Aviv 6997801, Israel}
\affiliation{Center for Physics and Chemistry of Living Systems,
  Tel Aviv University, Tel Aviv 6997801, Israel}

\author{Rudolf Podgornik}\thanks{Deceased, Dec. 28, 2024}
\affiliation{School of Physical Sciences and Kavli Institute for Theoretical Sciences, 
University of Chinese Academy of Sciences, Beijing 100049, China}
\affiliation{Wenzhou Institute, University of Chinese Academy of Sciences, Wenzhou, Zhejiang 325000, China}
\affiliation{CAS Key Laboratory of Soft Matter Physics, Institute of Physics, 
Chinese Academy of Sciences, Beijing 100190, China}

\author{David Andelman}
\affiliation{School of Physics and Astronomy,
  Tel Aviv University, Tel Aviv 6997801, Israel}
\email{andelman@post.tau.ac.il}
\affiliation{Center for Physics and Chemistry of Living Systems,
  Tel Aviv University, Tel Aviv 6997801, Israel}

\date{March 29, 2026}

\begin{abstract}
We study the effect of asymmetry in solvent and ionic size 
on the equilibrium properties of multivalent ionic solutions near a charged surface.
For a single ionic species in solution, we derive a generalized Grahame
equation at the charged surface. For general size ratio between the ions and the solvent,  
we obtain analytical results for the concentration profiles as a function of the distance from the surface.
For weak surface charge and small ion-to-solvent size ratio, 
the profile follows the classical Poisson-Boltzmann equation in dilute solution conditions.
However, for high surface charge and large ionic size, the concentration
profile saturates near the surface, leading to distinctive
dependencies of the solution properties on the surface charge density and
size asymmetry. Furthermore, the crossover between dilute and saturated 
regimes depends on the surface charge and ionic size asymmetry. We
suggest that a solution containing multiple ionic species of different
valencies and sizes stratifies close to the surface in the saturation regime. This leads to the formation
of layers that are ordered according to the ions' valency-to-size ratio.

\end{abstract}
\maketitle

\section{Introduction}
Electrostatic interactions between small and macro ions in aqueous
solutions govern the self-organization and dynamics of numerous soft
matter and biological systems, such as charged colloid suspensions,
charged bio-membranes, proteins, polyelectrolytes, and
DNA. For dilute ionic solutions, the Poisson-Boltzmann (PB) 
theory~\cite{Andelman95,Markovich18} is rather successful in describing 
equilibrium solution properties. The main advantage of the PB theory is its simplicity, which allows for analytical
or numerical solutions, in broad agreement with experiments.

At higher concentrations (above 1\,M for monovalent salts, corresponding
to an inter-ion distance of about 1.2\,nm), steric, and other short-range interactions 
as well as ion-ion correlations become important, 
and more refined theories are needed.
In particular, as the PB theory does not account for the finite ion size, it does not 
predict a maximal density (saturation) at charged surfaces, in cases where the ionic density approaches close packing.  
These effects require modifications of the PB theory, either in terms of
corrections that go beyond mean field, non-electrostatic contributions,
or using a lattice-gas model that accounts 
for the mixing entropy more correctly~\cite{Markovich18,Blossey2017}.

One of these extensions is the {\it sterically-modified} PB (sMPB) theory 
that accounts for the finite ionic and solvent
(water) molecular volume via a lattice-gas model for the mixing entropy~\cite{Borukhov1997,Borukhov2000}. 
The conceptual basis of this approximation is to combine the Poisson
equation of electrostatics with a lattice-gas entropy of mixing accounting  
also for the solvent entropy.  

In the more specific case where the ions and
solvent (water) molecules have roughly the same size, 
it is possible to obtain~\cite{Markovich18,Borukhov1997,Borukhov2000} 
analytical predictions and a description of the ion 
concentration saturation close to charged surfaces.
The more general case of ions and solvent having a different molecular size has also been
studied in experiments, theory, 
and numerical simulations~\cite{wen,zhou1,zhou2,gupta,avraham,suss,han,huang,podgornik,kumari}. 
For example, ionic solutions containing multiple counterions were shown to 
exhibit stratification close to a strongly charged surface, 
where the concentration of different counterion species peaks
at different distances from the surface. 

In the present work, we
incorporate size asymmetry effects using the
Flory-Huggins entropy of mixing. We extend the mean-field theory used in
Refs.~\cite{podgornik,kumari}, and present results for the
multivalent ion concentration and electrostatic potential profiles close to charged surfaces.
One of our main results is to show how the ionic valency-to-size ratio 
determines the order in which different ionic layers stratify close to the surface.

The outline of this paper is as follows. Section~II presents the
model applicable to a single counterion species near a charged surface, 
whose size differs from the solvent's. A modified PB equation is obtained in the dilute and saturated
regimes, and the effect of size-asymmetric steric repulsion is presented for strongly charged surfaces.
Then, in Sec.~III, we generalize the model for a multi-species ionic solution, and the profile equations, 
the Grahame equation, and the osmotic pressure are derived. 
We apply our findings from the single-counterion case to propose a theoretical argument for the stratification 
effect in multi-species ionic solutions. We end up by presenting the conclusions and 
connection with experiments in Sec.~V.

\section{Model for a single counterion solution}
\label{sec2}
We derive the free energy of an aqueous ionic solution 
using a discrete lattice-gas formulation with the ions and solvent
molecules fully occupying a three-dimensional cubic lattice. The lattice
cell volume, $a^3$, is taken as the volume of a solvent molecule.

In the first step, we consider a solution containing only  
a single ionic species of molecular volume $a^3v$ and charge
$q\,{=}\,ze<0$ where $z$ is the valency and $e$ the electronic charge. 
The parameter $v$ is the ratio between the
molecular volume of the counterion and that of the solvent ($a^3$).

The free energy, $F=U_{\rm el}-TS$, is expressed in terms of the local
electrostatic potential $\psi(\vecr)$ and the counterion number concentration $c(\vecr)$ 
and volume fraction $\phi(\vecr)=a^3vc(\vecr)$. The electrostatic contribution is~\cite{Markovich18}
\be
    U_{\rm{el}} = \int \text{d}^3{r} 
    \left( -\frac{\varepsilon}{2}\left|\nabla\psi\right|^{2} \ + \ qc\psi\right). 
\ee
The first term is the electric field's self-energy, where
$\varepsilon$ is the dielectric constant; the second term is the sum of
the ions' electrostatic energy. The Flory-Huggins entropic contribution to the free energy is written as
\be
    -TS = \frac{\kB T}{a^3}\int\text{d}^3{r}\left(\frac{\phi}{v}\ln\phi \ + \ 
    \left(1-\phi \right) \ln \left( 1-\phi \right) \right),
\ee
where $\kB T$ is the thermal energy.  The first term is
the counterions mixing entropy.
The second one is the mixing entropy
of the solvent whose volume fraction is $1-\phi$.

At equilibrium, the system has a spatially uniform thermodynamic pressure $P$, which
can be derived, {\it e.g.,} through the contact theorem~\cite{holovko,benyaakov}. 
It yields the following equation of state,
\be
\label{P_eq_single}
\begin{split}
    -\frac{Pa^3}{k_BT}=\frac{a^3\varepsilon}{2\kB T} |\nabla\psi|^2 + \ln \left(1-\phi \right) +w\phi = \rm{const} \ .
\end{split}  
\ee
where we make use of another size parameter, $w \equiv 1-1/v$. 
The first term on the right-hand-side of Eq.~(\ref{P_eq_single})
is the contribution to the pressure from the electrostatic stress, while
the second and third terms are coming from the lattice-gas entropy
pressure. At equilibrium, the total pressure is uniform. It can be evaluated at a reference position,
where the electrostatic field vanishes. This determines the constant of Eq.~(\ref{P_eq_single}).

\subsection{Counterion profile equations}
Minimizing the total free energy with respect to $\psi$ 
yields the Poisson equation, expressed in terms of volume fraction,
\be
\label{poisson_eq_single}
    \nabla^{2}\psi=-\frac{1}{\varepsilon}\frac{q\phi(\vecr)}{a^3v} \ ,
\ee
Next, minimizing the free energy with respect to $c$ (or $\phi$) gives 
the equation of state for the ionic concentration,
\be
\label{EoS_c(u)_single}
  \frac{\phi}{\left(1-\phi\right)^{v}}=\frac{\phi_{\rm{ref}}}
     {\left(1-\phi_{\rm{ref}}\right)^v}\,{\rm e}^{-\beta q\psi} \ .
\ee
where 
$\beta=1/\kB T$ is the inverse thermal energy, and $\phi_{\rm ref}$ and $1-\phi_{\rm ref}$
are, respectively, the counterion and solvent volume fractions at the reference position, 
where $\psi_{\rm ref}=0$.
Equation (\ref{EoS_c(u)_single}) 
for general $v$ is a transcendental equation with no analytical solution for $\phi$. 
Together with Eq.~(\ref{P_eq_single}), these two equations can be solved 
with the proper boundary conditions in
order to obtain the density profile $\phi(\vecr)$
and local potential $\psi(\vecr)$.

We can now derive the
profile equations in the dilute ($\phi\ll 1$) and saturation
($1-\phi \ll 1$) limits. In the following, we define $\eta$ as the solvent local fraction,
\be
\eta(\vecr)\equiv 1-\phi(\vecr).
\ee
For the dilute limit, neglecting the solvent contribution in
Eq.~(\ref{EoS_c(u)_single}), we recover the standard Boltzmann
distribution,
\be
\label{Boltzmann}
  \phi(\vecr) \approx \phi_{\rm{ref}}{\rm e}^{-{\beta}q\psi(\vecr)} \ .
\ee
On the other hand, at the saturation limit, the solvent volume fraction
$\eta$ is the small parameter. We rewrite
Eq.~(\ref{EoS_c(u)_single}) as
\be
\label{Eq7}
\begin{split}
  \frac{\eta(\vecr)}{\left[1-\eta(\vecr)\right]^{1/v}}={\rm e}^{-\beta[\mu - q\psi(\vecr)]/v} \ ,\\
 \\
 \beta  \mu =  \ln\phi_{\rm ref} - v\ln(1-\phi_{\rm ref}) \ .
\end{split}
\ee
Expanding to first order in $\eta$, we find that the solvent volume
fraction follows a modified Boltzmann distribution,
\be    
\label{Eq8}
    \eta(\vecr) \approx {\rm e}^{-\beta[\mu - q\psi(\vecr)]/v} \ ,
\ee
while the counterion volume fraction approaches saturation
with an exponentially small correction term that depends on the electrostatic potential,
\be
\label{single_sat_approx}
    \phi(\vecr) \approx 1- {\rm e}^{-\beta\left[\mu - q\psi(\vecr)\right]/v} \ .   
\ee

Compared to  Eq.~(\ref{Boltzmann}) of the dilute limit, 
Eqs.~(\ref{Eq7})--(\ref{single_sat_approx}) of the saturation  limit
contain an extra factor of $1/v$ in the exponent.
A simple argument can explain the origin of this size dependency. 
Consider that the ions in the saturated regime 
($\phi \,{\lesssim}\,1$) are in a system of volume $V$. 
Then, slightly dilute the system by adding a small amount of solvent, such that 
$\eta \to \eta + \delta\eta$. As all lattice cells are occupied, 
the ionic volume fraction changes by
$\delta \phi=-\delta \eta$, and each solvent molecule
replaces $1/v$ ions. The resulting change in free energy has three
contributions:
\bea 
& \left(\frac{V}{a^3}\right) \kB T\left( \ln\eta \ - \  1\right)\, \delta \eta & \textrm{mixing~entropy~gain} 
,\nonumber \\ 
& & \nonumber\\
& \left(\frac{V}{v a^3}\right)  \,q\psi \, \delta\phi = -\left(\frac{V}{v a^3}\right) \,q\psi \,\delta\eta  & \textrm{electrostatic~penalty}
 , \nonumber \\ 
 & & \nonumber \\
& \left(\frac{V}{v a^3}\right) \mu\,  \delta\phi= -\left(\frac{V}{v a^3}\right)\mu\, \delta\eta  & \textrm{chemical-potential} 
.\nonumber\\
\eea
Minimizing the total change in free energy with respect to 
$\delta\eta$, we obtain $\delta\eta = \exp\left[ -\beta (\mu - q\psi)/v \right]$, in agreement with Eq.~(\ref{Eq8})

\subsection{Counterion adsorption on a charged surface}
\label{sec_surface}
We now explore the ionic profile for the single counterion species near a charged
surface with charge density $\sigma>0$. We assume that the system is
homogeneous in the directions parallel to the surface. The 
profiles are then only a function of the distance away from the surface, $x$, making the
system one-dimensional (1D). 
As the pressure is uniform, it can be evaluated [ Eq.~(\ref{P_eq_single})] at infinite distance from the surface, 
where the electrostatic field vanishes $\psi'(x{\rightarrow}\infty){=}0$. 
The counterions have to balance the surface charge, and their concentration vanishes at infinity, 
$\phi(x{\rightarrow}\infty){=}0$ and $\eta(\infty){=}1$. Hence, the pressure in this counterion-only case is zero, and Eq.~(\ref{P_eq_single}) 
can be rearranged as
\be
\label{Pel=Pos}
    \ln \eta +w\phi=-\frac{a^3\varepsilon}{2\kB T} |\nabla\psi|^2 .
\ee

\subsubsection{The generalized Grahame equation}
\label{sec_phis}
Substituting the electrostatic boundary
condition, $\psi'(x{=}0){=}-\sigma/\varepsilon$ in Eq.~(\ref{P_eq_single}), we obtain
the condition for the counterion fraction at the surface, $\phis\equiv\phi(0)$, 
expressed via a dimensionless surface parameter $\zeta$,
\be
\label{phis_zeta}
  \ln\left(1 - \phis\right) + w \phis = -\zeta \ .
\ee
This is a modification of the Grahame equation~\cite{Markovich18} where 
\be
\label{zeta_def}
\zeta = 2\pi \lB a^3\left(\frac{\sigma}{e}\right)^{2}=\frac{a^3}{2\pi\lB l_{\rm GC}^2} \ ,
\ee
and the Bjerrum length $\lB$ and Gouy-Chapman length $l_{\rm GC}$~\cite{Markovich18}
are defined as usual,
\be
  \lB = \frac{e^2}{\varepsilon\kB T},\ \ \
  l_{\rm GC} = \frac{1}{2\pi\lB}\frac{e}{\sigma} \ .
\ee
Since the dimensionless surface parameter $\zeta$ does not depend on $v$, the
effect of the ion size on $\phis$ enters
only through the parameter $w=1-1/v$ defined above.

\begin{figure}
\begin{centering}
\includegraphics[width=0.4\textwidth]{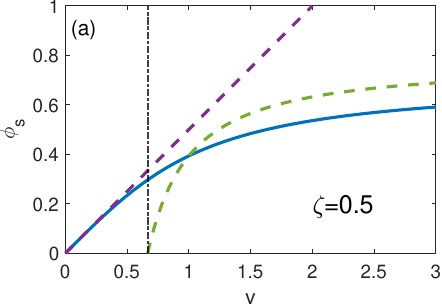}
\includegraphics[width=0.4\textwidth]{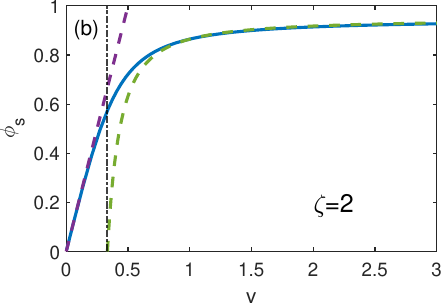}
\end{centering}
\caption{Counterion fraction at the charged surface $\phis$
(solid blue line) as a function of the size ratio $v$ [Eq.~(\ref{phis_W})].  
(a) The weakly charged surface case; and (b)
the strongly charged surface case. 
Convergence to the low $v$ limit (dashed purple line), 
Eq.~(\ref{grahame_PB}), is seen in both cases,
whereas in the high $v$ limit (dashed green line),
Eq.~(\ref{W_approx}) is a good approximation only for strongly charged surfaces
where $\phis$ is close to saturation. 
The vertical dash-dotted black line marks the crossover between the two limiting behaviors,
$v^{*}=1/(1+\zeta)$.
 }
\label{fig1}
\end{figure}

The solution to the transcendental Eq.~(\ref{phis_zeta}) can be
expressed in terms of the Lambert $W$ function (also called the {\it product logarithm}),
$W(u)$, defined as the inverse function of $u(W)=W {\rm e}^{W}$,
\be
\label{phis_W}
    \phis = 1+\frac{1}{w} W\left(-w{\rm e}^{-\zeta- w} \right) \ .
\ee
Equation~(\ref{phis_W}) is one of our central results. It can be
viewed as a generalized Grahame equation, relating the counterion
fraction at the surface, $\phi(0)=\phis$, with the surface charge density $\sigma$ for
a single counterion species of arbitrary size asymmetry. 

Equation~(\ref{phis_W}) is presented in Fig.~\ref{fig1} for both a weakly charged surface ($\zeta=0.5$) 
and a strongly charged one ($\zeta=2$).
In addition, we present three limiting cases. First, substituting $v=1$ ($w=0$, no asymmetry) 
reproduces the corresponding Grahame equation for the symmetric MPB case~\cite{Markovich18},
\be
\phis=1-{\rm e}^{-\zeta} \ .
\ee 

The second limit corresponds to saturation of the ionic concentration near the surface. 
It can be treated by assuming a large surface charge density $\sigma$ so that $\zeta$ is
sufficiently large, and the argument of $W$ is exponentially small. Then,
we can expand $W(u)$ to first order in $u$, 
$W(u\,{\ll}\, 1)\approx u$, yielding
\be
\label{W_approx}
    \phi_{\rm s}\approx1-{\rm e}^{-\zeta-w}.
\ee

The argument of $W$ in Eq.~(\ref{phis_W}) depends not only on $\zeta$,
but also on $v$ (through $w=1-1/v$). The third limit is that of small ionic size $v$.
Even for large $\zeta$, if the ionic size is
sufficiently small such that $v\lesssim v^* = 1/(1+\zeta)$, we cannot 
use the expansion of $W$ for small arguments. Instead, we should use the asymptotic
expansion for large arguments, $W(u\gg 1)\sim \ln u$. This is the case
in the limit of point-like ions ($v\rightarrow 0$), holding for a dilute
solution. Then, we recover the Grahame equation of the PB model~\cite{Markovich18},
\be
\label{grahame_PB}
  c_{\rm s}\approx 2\pi \lB\left(\frac{\sigma}{e}\right)^{2} \ ,
\ \ \ \phis = a^3 v c_{\rm s} \ .
\ee
Thus, $\phis$ in this limit is proportional to $v$.

In Fig.~\ref{fig1}, we plot $\phis$ as a
function of the size asymmetry $v$, 
for small $\zeta$ [$\zeta{=}\,0.5$ in \ref{fig1}(a)] and large $\zeta$ [$\zeta{=}\,2$ in \ref{fig1}(b)]. 
The solid blue line represents the full solution from Eq.~(\ref{phis_W}). 
For large $\zeta$ in \ref{fig1}(b), the approximate
Eq.~(\ref{W_approx}) (dashed green line) converges well to the full solution in the limit
of large $v$, reaching the maximum value of
$1-{\rm e}^{-\zeta-1}$. However, the approximation 
substantially deviates from the full solution for small $\zeta$ [dashed green line in Fig.~\ref{fig1}(a)].
The crossover asymmetry $v^*$ is marked by a vertical dash-dotted line. 
For $v<v^*$, the solution for $\phis$ tends to the
linear curve $ v$, given by the Grahame equation,
Eq.~(\ref{grahame_PB}) (dashed purple line). This is clearly evident for both values of $\zeta < 1$ and $ \zeta> 1$ 
shown in the figure.

\subsubsection{Ionic profile: $\phi(x)$}
Taking the spatial derivative of Eq.~(\ref{P_eq_single}), and combining it
with the Poisson equation [Eq.~(\ref{poisson_eq_single})], we get a
 first-order differential equation relating $\phi(x)$ and its derivative $\phi'(x)$. This
equation can be integrated, yielding
\bea
\label{single_exact_solution}
x(\phi)= &&\frac{1}{\alpha}\sqrt{\frac{a^3}{8\pi l_{\rm B}}} \times\\
&&\intop_{\phis}^{\phi(x)}{\rm d}\phi\ 
    \frac{(1-\phi)^{-1}-w}{\phi\sqrt{-\ln\left(1-\phi\right)-w\phi }} \ .
    \nonumber
\eea
where $\alpha=|z|/v$ is the valency-to-size ratio.  
Once the boundary value $\phis$ is obtained from Eq.~(\ref{phis_W}), 
we can calculate numerically the profile $\phi(x)$ by inverting $x(\phi)$
from Eq.~(\ref{single_exact_solution}). Then, using
Eq.~(\ref{EoS_c(u)_single}) with $\phi_{\rm{ref}}=\phis$, the profile of the electrostatic potential $\psi(x)$
can be obtained as well.

In a saturated region, the right-hand-side of Eq.~(\ref{poisson_eq_single})
tends to a constant, $\psi'' \approx -q/(\varepsilon v a^3)$, 
and this 2nd-order ODE for 1D symmetry can be integrated twice. 
The resulting spatial dependence of the electrostatic potential is
parabolic~\cite{Borukhov1997},
\bea
\label{psi_lstar}
\beta e\psi(x) &\approx& -\frac{\zeta}{\alpha}+\frac{2\pi \lB\alpha}{a^3}\left(x-\ell^{*}\right)^{2}\ , \nonumber\\
\ell^{*}&=& a^3\sigma/\alpha e,
\eea
where the potential at the surface is set as the reference potential, $\psi_{\rm{ref}}=\psi(x{=}0)=0$. 
Clearly, there is an extremum at a distance $\ell^{*}$ from the surface.  
Since the counterion concentration profile depends exponentially on $\psi$, Eq.~(\ref{single_sat_approx}), 
the length $\ell^{*}$ characterizes well the thickness of the saturation layer.

\begin{figure}
\begin{centering}
\includegraphics[width=0.45\textwidth]{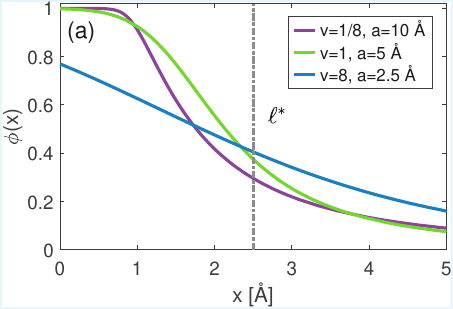}
\includegraphics[width=0.45\textwidth]{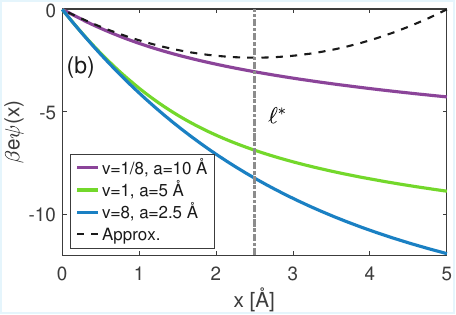}
\end{centering}
\caption{(a) Counterion volume fraction $\phi(x)=a^3v c(x)$, and (b) dimensionless electrostatic potential 
$\beta e\psi(x)$,  as a function of distance from the surface, $x$, for different solvent sizes and size asymmetries.
 In all cases, the ionic size is held constant, $a^3v = 125$\,\AA$^3$, the ions are monovalent, $z=1$, and $\alpha=|z|/v=1/v$. The
  surface charge density is $\sigma{=}2e/{\rm nm^{2}}$. The dashed-dotted
  vertical line marks the value of $\ell^{*} = a^3\sigma /\alpha e=2.5\,$\AA. 
  The approximate parabolic profile of the electrostatic potential (dashed black line) 
  for $v=1/8$ and $a=10$\,\AA\ is drawn in (b) from Eq.~(\ref{psi_lstar}).
  }
\label{fig2}
\end{figure}

Figure~\ref{fig2}a shows the counterion volume fraction profile, $\phi(x)$, 
as a function of the distance from the surface, $x$,
obtained from Eq.~(\ref{single_exact_solution}). 
Similarly, Fig.~\ref{fig2}b shows the respective electrostatic potential profile, obtained from Eq.~(\ref{Eq7}). The profiles shown
are for the symmetric case ($v{=}1$; green line), as well as for 
smaller size ratio between counterion and solvent
($v{=}1/8$; purple line) and a larger one
($v{=}8$; blue line). The solvent size $a$ is
adjusted while the counterion size $va^3$ is kept constant.
Since $a^3/\alpha \sim va^3$, the characteristic length $\ell^{*}$ is also unchanged. 

Figure 2 demonstrates
the enhanced saturation effect caused by larger solvent
molecules. This is due to the smaller entropy gained by mixing the
counterions with the larger solvent molecules. From a more intuitive
perspective, the large solvent molecules osmotically push the smaller
ions closer to the surface.

The volume fraction profiles intersect around $x\approx \ell^{*}$, demonstrating the
role of $\ell^{*}$ as a characteristic thickness of the saturated layer. 
For $x>\ell^{*}$, the parabolic approximation of the potential profile [Eq.~(\ref{psi_lstar})] 
is no longer valid and deviates largely from the exact profile. 
Clearly, this approximation can be used only for $x<\ell^{*}$

The length $\ell^*$ can be heuristically inferred from the
typical thickness of a saturated layer required to neutralize the
charged surface, as follows. The area per surface charge is
$e/\sigma$.  A single counterion of valency $z$ participating in the
saturated layer has volume $a^3v$ and neutralizes $|z|$ monovalent
surface charges, {\it i.e.}, an area of $|z|e/\sigma$. Placing the counterion inside a
fictitious rectangular box of base area $|z|e/\sigma$ and volume $a^3v$
leads to a box length of $va^3\sigma/|z|e$, which is exactly $\ell^{*}$ as derived in Eq.~(\ref{psi_lstar}).

\section{Free-energy and profiles for a multi-species ionic solution}
\label{sec3}
We now consider the generalization of the single ionic species presented in Sec.~\ref{sec2}
to the case of $M$ ionic species. Each species $i=1,...,M$, is
characterized by a molecular volume $a^3v_i$ and ionic charge $q_{i}=z_i e$
(valency $z_i$). The parameter $v_i$ is the ratio between the
molecular volume of the $i$-th ionic species and that of the solvent ($a^3$).

As in Sec.~\ref{sec2}, the free energy, $F=U_{\rm el}-TS$, is expressed in terms of the local
electrostatic potential $\psi(\vecr)$ and the ionic concentrations $c_i(\vecr)$ or, equivalently, volume fractions
$\phi_{i}(\vecr)=a^3v_ic_{i}(\vecr)$. The electrostatic ution is
\be
    U_{\rm{el}} = \int \text{d}^3{r} 
    \left( -\frac{\varepsilon}{2}\left|\nabla\psi\right|^{2}+\sum_{i=1}^{M} q_{i}c_{i}\psi\right). 
\ee
The first term is the electric field's self-energy, and the second term is the sum of
the ions' electrostatic energy. Then, the Flory-Huggins entropic contribution to the free energy is written as
\be
    -TS = \frac{\kB T}{a^3}\int\text{d}^3{r}\left(\sum_{i=1}^{M}
    \frac{\phi_{i}}{v_i}\ln\phi_{i}+
    \eta\ln\eta\right) \ .
\ee
The first term is the
mixing entropy of the ions summed over all ionic species.  
The second one is the mixing entropy
of the solvent. Assuming incompressibility, the solvent local volume fraction is
\be
\eta(\vecr)=1-\sum_{i=1}^{M}\phi_{i}(\vecr).
\ee

At equilibrium, the system has a spatially uniform thermodynamic pressure, which
can be derived through the contact theorem~\cite{holovko,benyaakov}. 
It yields the following equation of state that is a generalization of Eq. (\ref{P_eq_single}),
\be
\label{P_eq_general}
\begin{split}
    -\frac{Pa^3}{k_BT}=\frac{a^3\varepsilon}{2\kB T} |\nabla\psi|^2 + \ln \eta +\sum_{j=1}^M w_j \phi_{j}
    =\rm{const.}
\end{split}  
\ee
where the parameter $w_j$ is defined as 
\be
\label{wj}
w_j \equiv 1-1/v_j\ .
\ee  
The first term on the right-hand-side of Eq.~(\ref{P_eq_general})
is the contribution to the pressure from the electrostatic stress, 
while the second and third terms come from the lattice-gas entropy.

\subsection{Profile equations for an $M$  ionic species}
Minimizing the total free energy with respect to $\psi$ and $\{c_{i}\}$ (or $\{\phi_{i}\}$), $i=1,..., M$,
yields the Poisson equation, expressed in terms of volume fractions,
\be
\label{poisson_eq}
    \nabla^{2}\psi=-\frac{1}{\varepsilon}\sum_{i=1}^{M}\frac{q_{i}\phi_{i}(\vecr)}{a^3v_i} \ ,
\ee
and the equation of state for the ionic concentrations, 
\be
\label{EoS_c(u)_general}
    \frac{\phi_{i}}{\left(1-\sum_{i=1}^{M}\phi_{i}\right)^{v_i}}=\frac{\phi_{i}^{\rm {ref}}}
    {\left(\eta_{\rm{ref}}\right)^{v_i}}{\rm e}^{-\beta q_{i}\psi} \ ,
\ee
where $\phi_{i}^{{\rm{ref}}}$ and $\eta_{\rm{ref}}$ are, respectively, 
the volume fractions of the $i$-th species and the solvent at the reference position, where $\psi_{\rm{ref}}=0$.

Equation (\ref{EoS_c(u)_general}) represents $M$ algebraic equations, 
containing a polynomial of degree $v_i$. Together with Eq.~(\ref{P_eq_general}) we
have $M+1$ equations for the $M$ density profiles $\phi_i(\vecr)$
and for the local electrostatic potential $\psi(\vecr)$.

\subsection{Stratification of multi-counterion adsorption}

\begin{figure}
\begin{centering} 
\includegraphics[width=0.4\textwidth]{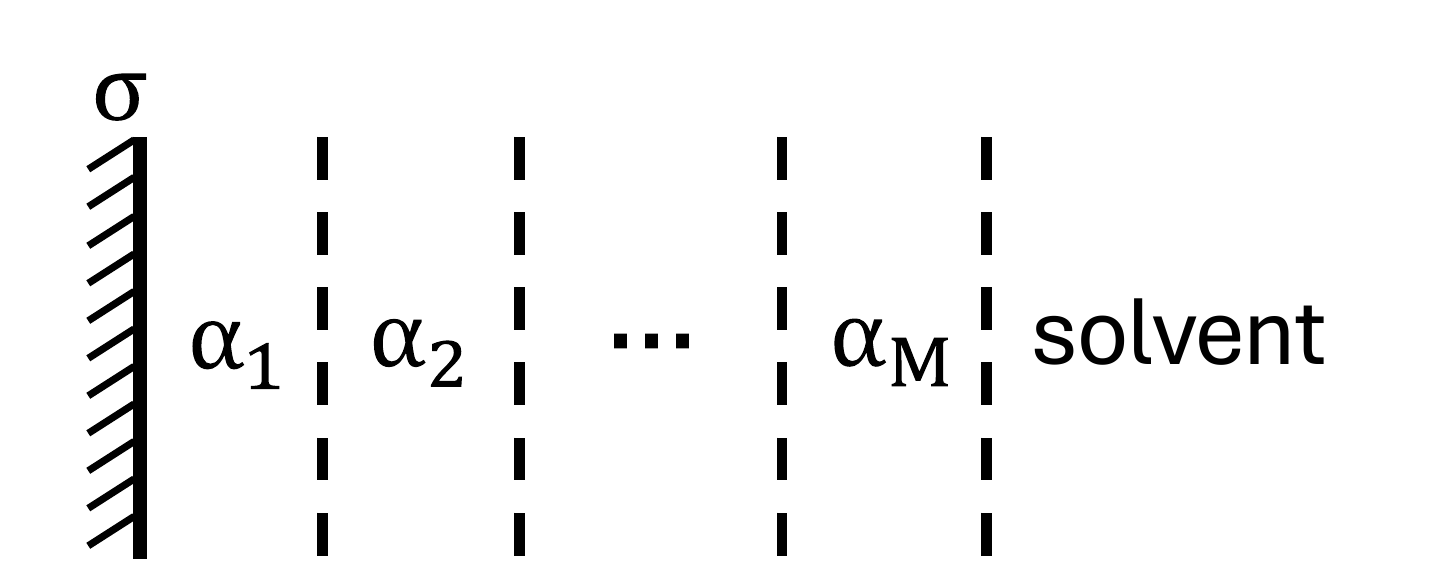} \par\end{centering}
\caption{{\small{}
        Schematic of a stratified counterion system near a charged surface. The label $\alpha_k$ 
        represents the dominating species in the layer $k$ away from the charged surface. And the stratified layers are 
        characterized by $\alpha_1>\alpha_2>\cdots>\alpha_M$.}     
                 }
\label{fig3}
\end{figure}

Consider a solution containing several counterion species of different size and valency. Close to a strongly
charged surface, some experiments~\cite{avraham,suss} and numerical works~\cite{zhou1,zhou2,wen} 
have found that the concentration of lower-valency counterions is higher near the surface.
In general, charged surfaces preferentially adsorb high-valency counterions. 
However, if the lower-valency ions are smaller, this preference can be reversed.
Moreover, the above-mentioned studies
suggest that the valency-to-size ratio, $\alpha= |z|/v$,
determines the spatial order of adsorption (stratification) from a multi-counterion solution \,---\,
the larger the value of $\alpha$, the closer the maximal concentration of
the ion species to the charged surface. The dependence on $\alpha$
is already seen, indeed, in the expressions for the electrostatic
potential $\psi$ and characteristic length $\ell^{*}$ of the saturated
region, Eq.~(\ref{psi_lstar}). We now give a semi-quantitative
argument to account for these observations.

The solution has $M$ negative ion species with
valencies $z_i$ and relative specific volumes $v_i$ (with respect to the solvent molecular size $a^3$), 
near a positively charged
surface. We assume that the surface is strongly charged, forming a
saturated layer of adsorbed counterions, as depicted in Fig.~\ref{fig3}. The saturation implies that
the electrostatic energy dominates the free energy of adsorption, 
and mixing entropy can be neglected. Let us further assume that the
counterion species are stratified into layers indexed by $k$
according to their distance from the surface. We examine the energy
cost of exchanging $n_k$ ions of layer $k$ with $n_{k+1}$ ions of
layer $k+1$.  For the exchange to be
energetically unfavorable, we will prove that the condition has to be $\alpha_{k+1}<\alpha_k$.

Taking the electrostatic potential in the two layers to be $\psi_k$
and $\psi_{k+1}$, the electrostatic energy cost of the exchange is
\bea
   \frac{1}{e} \Delta U_{\rm el}&=& \psi_k\left( n_{k+1}z_{k+1} - n_kz_k \right) \nonumber\\
   &&~+ ~\psi_{k+1}\left( n_kz_k - n_{k+1}z_{k+1} \right) \nonumber\\
   &=&\left( \psi_k-\psi_{k+1}\right) \left( n_{k+1}z_{k+1} - n_kz_k \right) \nonumber\\
   &=& -  |\psi_k-\psi_{k+1}| \left( n_{k+1}|z_{k+1}| - n_k|z_k| \right). \nonumber\\
\label{DUel}
\eea
Now, to maintain the condition of saturation in the $k$-th layer,
\be
   n_{k+1}v_{k+1} - n_kv_k = 0\ \ \Rightarrow \ \ n_k/n_{k+1} = v_{k+1}/v_k.
\ee
Substituting the above in Eq.~(\ref{DUel}) yields
\be
 \frac{ 1}{e}\Delta U_{\rm el} =  |\psi_k-\psi_{k+1}| n_{k+1} v_{k+1}
  \left( \alpha_k - \alpha_{k+1} \right),
\ee
where we recall that $\alpha_i=|z_i|/v_i$. Thus, the condition for having $\Delta
U_{\rm el} > 0$ (unfavorable exchange) is $\alpha_{k+1} <
\alpha_k$. This argument is quite general, provided the
electrostatic potential is sufficiently large.
Thus, with increasing distance from the surface, the layers are arranged in decreasing order of the valency-to-size ratio $\alpha$.

\section{Conclusions}

\begin{figure}
\begin{centering} 
\includegraphics[width=0.4\textwidth]{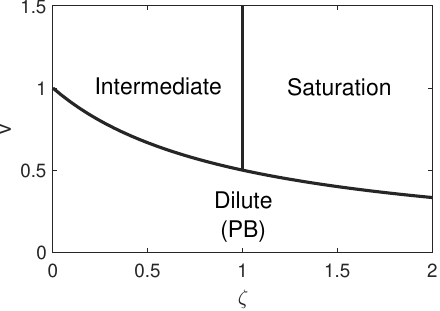} \par\end{centering}
\caption{{\small{}
        The three regimes of counterion concentration profiles, plotted on
    the plane of surface charge ($\zeta\sim \sigma^2$) and size asymmetry ($v$). 
    The crossover between the intermediate and saturation regimes occurs at $\zeta=1$, 
    and the one between the dilute and the two other regimes at  $v^{*}=1/(1+\zeta)$.}
            }
\label{fig4}
\end{figure}

We have studied the effects of the size asymmetry of ions and solvent
on the ionic concentration profiles in an electrolyte solution in contact with a charged
surface. We have obtained analytical results for the ionic volume fraction
adjacent to the surface, Eq.~(\ref{phis_W}), and (up to an integration)
for the full concentration profiles,
Eq.~(\ref{single_exact_solution}). 

The main findings are summarized in the diagram presented in
Fig.~\ref{fig4}. The behavior can be separated into three
qualitative regimes. In the dilute regime, the PB
theory predictions are valid. This regime can be reached by either decreasing the
surface charge ($\sigma$) or decreasing the ion size relative to the
solvent (the $v$ parameter). We have been concerned mainly with the opposite case, of a
solution whose concentration profile near the surface is
saturated. The saturation regime is reached by a high surface charge $\sigma$
and large ion size $v$ relative to the solvent. In this case, the
layer structure near the surface has distinctive dependencies on surface
charge and size asymmetry, as given by Eq.~(\ref{W_approx}). 

The line $v^*=1/(1+\zeta)$ marks the crossover between the PB regime and 
two regimes where the PB predictions are not valid anymore. For a weakly charged surface 
and large ion size ($\zeta <1$ and $v>v^*$), this intermediate regime 
is neither well described by the PB theory nor does it have a
saturated layer. The saturation regime occurs for $v>v^*$ and $\zeta>1$.

In addition, our model accounts for stratification in solutions containing multiple counterion species. 
In this case, separate saturation layers form near the charged surface. The model predicts that stratified layers
are ordered according to the valency-to-size ratio of the counterions, $\alpha=|z|/v$, 
in agreement with earlier observations~\cite{suss,avraham}.

Consideration of co-ion species is straightforward by applying the model presented in Sec. \ref{sec3}, as done previously by Maggs and Podgornik~\cite{podgornik}. 
For strongly charged surfaces, co-ions are depleted from the counterion-saturated layer by both the electrostatic repulsion and the steric hindrance. 
Therefore, we expect our results for the saturation limit to remain unaffected by the inclusion of co-ions.

Since we aim in this work to explore the ionic size effects, we have not
considered additional contributions such as specific ion-surface or
ion-ion interactions and polar effects. Furthermore, the assumption of a uniform
dielectric constant being equal to the solvent's $\varepsilon$ is
problematic in situations of counterion saturation. In these cases, the solvent molecules
are depleted from the region near the surface, or their orientation
are strongly affected by the ions~\cite{levy,iglic}. 
This implies that the present treatment overestimates the screening by the aqueous solvent 
and underestimates the electrostatic interactions. This observation is
supported by several models that investigated the combined effects of
medium polarizability and ionic finite size~\cite{gongadze,lopezgarcia,nakayama,JCP2018}.

The present work can be extended in several directions. Curved
geometries (e.g., a charged cylindrical surface or charged sphere) introduce another
length scale whose interplay with the ion size should be more complex to analyze
and may affect, for example, the Manning-Oosawa
condensation~\cite{manning,oosawa,manningoosawa} 
for the cylindrical case or the effective surface charge for the charged sphere. 
Another interesting direction is to consider a continuous
distribution of ion size and its effects on the 
concentration and charge density profiles of counterions near charged surfaces.

\bigskip\bigskip
\noindent{\bf Acknowledgments.}
This article is dedicated to the memory of our dear friend and colleague, 
Prof.~Rudolf (Rudi) Podgornik, who participated in the earlier stages of this work.
DA acknowledges partial support from the NSFC-ISF Research Program, jointly funded by the 
National Science Foundation China (NSFC) 
and the Israel Science Foundation (ISF) 
under grant No.~3396/19, and ISF grant No. 226/24. HD acknowledges
partial support from the NSFC-ISF (grant No.~3159/23) and the ISF (grant No.~1611/24).


\end{document}